\documentclass[pre,twocolumn]{revtex4-1}

\usepackage{dcolumn}
\usepackage{bm}

\usepackage[utf8]{inputenc}
\usepackage[T1]{fontenc}
\usepackage{mathptmx}

\usepackage{graphicx}
\usepackage{amsmath,amssymb}
\usepackage{xcolor}
\usepackage{chemformula}
\usepackage{ulem}
\usepackage{mhchem}

\usepackage{graphicx}
\usepackage{amsmath}
\usepackage{mathtools}
\usepackage{amssymb}
\usepackage{pst-node}

\usepackage{verbatim}   
\usepackage{color}      
\usepackage{subfigure}  
\usepackage{hyperref}   

\usepackage{braket} 

\newcommand{\bnabla}{\mbox{\boldmath $\nabla$}}

\newcommand{\ba}{\begin{eqnarray}}
\newcommand{\ea}{\end{eqnarray}}
\newcommand{\be}{\begin{equation}}
\newcommand{\ee}{\end{equation}}

\begin{document}

\title{Stationary distributions of propelled particles as a system with quenched disorder}

\author{Derek Frydel}
\affiliation{Department of Chemistry, 
Universidad Técnica Federico Santa María, Campus San Joaquin, Santiago, Chile}

\date{\today}

\begin{abstract}
This article is the exploration of the viewpoint within which propelled particles in a steady state are regarded
as a system with quenched disorder. The analogy is exact when the rate of the drift orientation vanishes and the
linear potential, representing the drift, becomes part of an external potential, resulting in the effective potential
$u_{eff}$. The stationary distribution is then calculated as a disorder-averaged quantity by considering all contributing
drift orientations. To extend this viewpoint to the case when a drift orientation evolves in time, we reformulate
the relevant Fokker-Planck equation as a self-consistent relation. One interesting aspect of this formulation is that
it is represented in terms of the Boltzmann factor $e^{-\beta u_{eff}}$.   In the case of a run-and-tumble model, 
the formulation reveals an effective interaction between particles.  
\end{abstract}

\pacs{
}

\maketitle


\section{Introduction}

Within the two standard models of propelled motion, the run-and-tumble (RTP) and active Brownian 
particle (ABP) model, particles are subject to a drift of constant magnitude v0 but randomized orientation. 
The time evolution of the drift is what prevents a system from attaining an equilibrium. The evolution of 
the orientation in each model is governed by a different stochastic process. In the RTP model, the new 
direction is assigned sporadically at intervals drawn from Poisson distribution. A new orientation can 
take on any value with equal probability. In the ABP model, the orientation undergoes diffusion. The rate 
of orientation change in the RTP model is $\alpha$ and the angular diffusion in the ABP model is $D_r$.

Despite the apparent simplicity of an ideal-gas model of propelled particles, there is no available analytical 
solution for stationary distributions. One noted exception is the RTP model in one dimension with drift limited 
to two values, $v=\pm v_0$ \cite{Schnitzer93,Cates08,Cates09,Angelani17,Dhar18,Dhar19,Razin20}.  
Yet even a simple extension to three drifts $v=0,\pm v_0$ leads to considerable increase in complexity 
\cite{Basu20}.  (The third model of propelled motion is the active Ornstein Uhlenbeck particles, 
 AOUP \cite{Martin21a,Martin21b}; however, in this work we exclusively focus on the RTP and ABP models.)

In this work, we take a different point of view to characterize stationary distributions of propelled particles. 
We start by considering a stationary state of propelled particles at exactly $\alpha=D_r=0$.  Under these conditions, 
the unit vector ${\bf u}_v$, representing orientation of a drift, stops evolving in time and as a consequence the system 
attains equilibrium. The result is a mixture of particles with different drift orientations. And because the drift 
orientations are randomly distributed, the situation corresponds to a system with quenched disorder. 
The stationary distribution is a disorder-averaged distribution that takes into account all drift orientations.

The resulting distribution for the condition $\alpha = Dr = 0$ represents the largest deviation from the distribution 
for the same system but for passive Brownian particles. Since in the limit  $\alpha\to\infty$ and/or $D_r\to\infty$
the distribution converges to that of passive particles, this limit is generally regarded as representing an equilibrium. 
The suggestion, therefore, that the opposite limit $\alpha,D_r\to 0$ corresponds to an equilibrium appears to 
contradict this view. If we look into the entropy production $\Pi$ that is used as a quantification of distance from 
the equilibrium, we find that $\Pi$ vanishes as $\alpha,D_r\to 0$, supporting the claim that this limit represents an 
equilibrium. The opposite limit $\alpha\to\infty$ is found to yield the largest possible value of $\Pi$, indicating the 
largest deviations from equilibrium --- a surprising result given that the distribution in that limit is the same as that for 
passive Brownian particles.

The central quantity that emerges in analyzing the limit $\alpha,D_r\to 0$, 
is the effective external potential, which is the original external potential $u_{ext}$ plus the linear 
potential representing a drift, $u_{eff} = u_{ext} + [{\bf u}_v\cdot{\bf r}] v_0/D$.  
One way to go beyond the decoupled limit, is to expand the stationary distribution perturbatively 
as $n \approx n_0 + \alpha n_1$. 
This approach, however, leads to a rather complex expression for $n_1$ without offering 
valuable insights.  Instead we reformulate the 
stationary Fokker-Planck equation (FP) as a self-consistent relation (SC).  The 
central quantity of the SC formulation is the Boltzmann factor $e^{-\beta u_{eff}}$.  
Within the SC formulation, propelled particles appear as if they were coupled, but the effective attraction has the 
“chemical” origin and arises when particles of different drift orientations are regarded as different species that 
undergo a continuous conversion. The SC formulation is used as a basis for numerical computation of stationary 
distributions, an alternative procedure to dynamic simulations.

This work is organized as follows.  In Sec. (\ref{sec:framework}) we introduce a general FP 
equation of propelled particles for an arbitrary dimension $d$.  In Sec. (\ref{sec:decoupled}) 
we consider exact distributions for a decoupled condition $\alpha=D_r=0$, which represents
the system with quenched disorder.  In Sec. (\ref{sec:SC}) we develop the self-consistent
framework for solving the stationary FP equation.  The goal of such a framework is to gain 
insights as well as to look for alternative numerical schemes other than dynamic simulations.  
 In Sec. (\ref{sec:diss}) we analyze the entropy production of a two-state RTP model.

\section{Theoretical framework}
\label{sec:framework}

The motion of an ideal-gas of propelled particles in a general $d$-dimensional space, with 
both RTP and ABP type of motion, is governed by the following Fokker-Planck equation (FP)
\ba
\frac{\partial n}{\partial t} &=& D\nabla^2 n - v_0 {\bf u}_v \cdot \bnabla n
+ \beta D\bnabla \cdot [n \bnabla u_{ext} ]  \nonumber\\
&-&\alpha \bigg[ n - \frac{1}{\Omega_v}\int d{\Omega}_v\, n \bigg] + D_r \hat L_{S} n, 
\label{eq:FP}
\ea
where the distribution $n\equiv n({\bf r},{\bf u}_v,t)$ is the function of the position ${\bf r}$, drift 
orientation ${\bf u}_v$ (${\bf u}_v$ is a unit vector), and time $t$, and is normalized to unity as 
$\int d{\bf r} \,n({\bf r},{\bf u}_v,t)=1$.  
The first line in Eq. (\ref{eq:FP}) governs the evolution of particle positions and involves 
standard diffusion, drift of constant magnitude $v_0$, and the interaction with external forces due 
to a conservative potential $u_{ext}({\bf r})$.   

The second line in Eq. (\ref{eq:FP}) governs the evolution of the unit vector ${\bf u}_v$ which 
determines the orientation of a drift.  The time evolution of ${\bf u}_v$ is what prevents the system 
from attaining equilibrium.  The first term gives rise to the RTP type of motion, where  
$\Omega_v=\int d{\Omega}_v$ is the area of a unit sphere in a given dimension.  
The RTP motion is represented as a "reaction" process where particles of different orientations 
are continuously created and destroyed yet their total number is conserved.  
The ABP motion is represented as a diffusion of a unit vector ${\bf u}_v$ on a surface of a sphere
and the operator $\hat L_S$ is a spherical Laplacian operator on the $(d-1)$-sphere.

For the explicit dimension $d=2$ the second line in Eq. (\ref{eq:FP}) becomes 
\be
\frac{\partial n}{\partial t} = -\bnabla \cdot {\bf j} 
-\alpha \bigg[ n - \frac{1}{2\pi}\int_{0}^{2\pi} d\theta_v\, n \bigg] + D_r \frac{\partial^2 n}{\partial\theta_v^2}. 
\label{eq:FP2D}
\ee
where we introduce the flux
$$
{\bf j} ({\bf r},{\bf u}_v) = -D\bnabla n + v_0 {\bf u}_v n - \beta D\bnabla \cdot [n \bnabla u_{ext}].  
$$


Because Eq. (\ref{eq:FP}) and Eq. (\ref{eq:FP2D}) involve creation-destruction of particles with 
different orientations, it is not immediately 
clear if the total number of particles is conserved.  To demonstrate that this is the case, we 
integrate Eq. (\ref{eq:FP2D}) over the space domain within which the system is confined, 
\be
\frac{\partial N}{\partial t} = -\alpha \bigg[ N - \frac{1}{\Omega_v}\int d{\Omega}_v\, N \bigg] 
+ D_r \frac{\partial^2 N}{\partial\theta_v^2}, 
\label{eq:FPJ1}
\ee
where we define the number of particles with particular orientation as $N({\bf u}_v) = \int d{\bf r}\,n({\bf r},{\bf u}_v)$.  
Note that $\int d{\bf r}\, \bnabla \cdot {\bf j}  = 0$ since particles do not enter or leave the prescribed domain.  
Finally, if we integrate Eq. (\ref{eq:FPJ1}) over all orientations and define $\bar N = \frac{1}{\Omega_v} \int d \Omega_v\,N({\bf u}_v)$, 
we have
\be
\frac{\partial \bar N}{\partial t} 
= -\alpha \big[ \bar N - \bar N \big] + D_r  \Big[\frac{\partial N}{\partial\theta_v}\Big]_{0}^{2\pi}
= 0,  
\ee
where the second term cancels out as a result of periodic boundary conditions.  The total 
number of particles, therefore, is conserved. 

At this point, we introduce the "effective" external potential defined as 
\be
\beta u_{eff} = \beta u_{ext} - \frac{v_0}{D}[{\bf u}_v\cdot{\bf r}], 
\label{eq:ueff}
\ee
which is the external potential plus the linear potential for representing a drift.  
As we limit our analysis to stationary distributions, the time-independent FP equation of 
interest is 
\be
0 = D\nabla^2 n 
+ \beta D\bnabla \cdot [n \bnabla u_{eff} ]  
-\alpha \bigg[ n - \frac{1}{\Omega_v}\int d{\Omega}_v\, n \bigg] + D_r \hat L_{S} n. 
\label{eq:FPS}
\ee
The stationary distribution that accounts for all orientations is defined as 
$$
\bar n({\bf r}) = \frac{1}{\Omega_v} \int d{\Omega}_v\, n({\bf r},{\bf u}_v),
$$
where the bar indicates the averaging procedure.





\section{Exact treatment in a decoupled limit, $\alpha= 0$ and $D_r=0$}
\label{sec:decoupled}

In this section we obtain distributions $n$ for a decoupled condition given by $\alpha=D_r=0$.  
Because under such circumstances ${\bf u}_v$ stops to evolve in time, 
the system is in equilibrium, but the distribution of drift orientations introduces quenched disorder.  


By setting both $\alpha$ and $D_r$ to zero, 
Eq. (\ref{eq:FPS}) reduces to 
\be
0 = D\nabla^2 n_0 + \beta D \bnabla \cdot [n_0 \bnabla u_{eff}]. 
\ee
The result is the standard diffusion equation for a particle in the external potential $u_{eff}$.  
The solution is proportional to the Boltzmann weight 
$$
n_0({\bf r},{\bf u}_v) \propto e^{-\beta u_{ext}} e^{[{\bf u}_v\cdot{\bf r}] v_0/D}.  
$$  
The subscript "0" is used to indicate that the solution is true only for the case $\alpha=D_r= 0$.
The actual stationary 
distribution is obtained by averaging over all possible drifts ${\bf u}_v$ uniformly 
distributed over all orientations and given by
\be
\bar n_0({\bf r}) \propto  \int d\Omega_v\, e^{-\beta u_{ext} +[{\bf r} \cdot{\bf u}_v]v_0/D}.  
\label{eq:n0_ave}
\ee
Quenched disorder is the inherent feature of the system in the decoupled limit.  

If $u_{ext}$ depends on particle positions only, then the Boltzmann factor can be separated
and above equation can be written as 
$$
\bar n_0({\bf r}) \propto  e^{-\beta u_{ext}({\bf r})}  \int d\Omega_v\, e^{[{\bf r} \cdot{\bf u}_v]v_0/D}. 
$$
All orientations in the above formulation are equally likely and there is no bias for any particular
direction.  But if the external potential contributes to particle orientations, $u_{ext}\equiv u_{ext}({\bf r},{\bf u}_v)$, 
a case that might arise for particles with dipole moment, the orientation would 
no longer be distributed uniformly and we would have 
$$
\bar n_0({\bf r}) \propto  
\int d\Omega_v\, e^{-\beta u_{ext}({\bf r},{\bf u}_v)} e^{[{\bf r} \cdot{\bf u}_v]v_0/D}.  
$$
Such orientation bias would reduce quenched disorder.  In this work, 
however, we limit our interest to the position dependent potentials.


\subsection{harmonic trap}

We next consider a number of specific potentials.  For a harmonic potential $\beta u_{ext} = \frac{1}{2} Kr^2$, 
$u_{eff} = \frac{1}{2} Kr^2 - [{\bf u}_v\cdot{\bf r}] v_0/D$ and  the Boltzmann distribution representing the 
decoupled limit is 
\be
n_0({\bf r},{\bf u}_v) \propto e^{-\frac{1}{2} \beta Kr^2} e^{[{\bf u}_v\cdot{\bf r}] v_0/D}.  
\label{eq:n0_K}
\ee
The disorder averaged distribution is obtained using Eq. (\ref{eq:n0_ave}).  For dimension $d=2$
we have $d\Omega_v=d\theta_v$, leading to 
$$
\bar n_0(r) \propto  e^{-\frac{1}{2} \beta Kr^2}  \int_0^{2\pi} d\theta_v\, e^{r\cos\theta_v v_0/D}.  
$$
After evaluating the integral we find 
\be
\bar n_0(r) =  
\bigg[\bigg(\frac{\beta K}{2\pi}\bigg) e^{-\frac{1}{2} \beta K r^2}\bigg]  
\bigg[e^{-\frac{v_0^2}{2D^2 \beta K}}  I_0\left(\frac{v_0r}{D}\right)\bigg].  
\label{eq:n02D}
\ee
The two terms in square-brackets indicate different contributions.  The first is the usual Gaussian 
distribution for passive particles in a harmonic potential.  The second term, represented by the modified 
Bessel function $I_0(x)$, is the contribution due to propelled motion.  
This term diverges far away from the center of the trap as $I_0(x) \approx e^{x}/\sqrt{2\pi x}$ and 
gives rise to particle deposition at the border of a trap \cite{rudi18}.

\begin{figure}[h] 
 \begin{center}
 \begin{tabular}{rrrr}
\includegraphics[height=0.21\textwidth,width=0.23\textwidth]{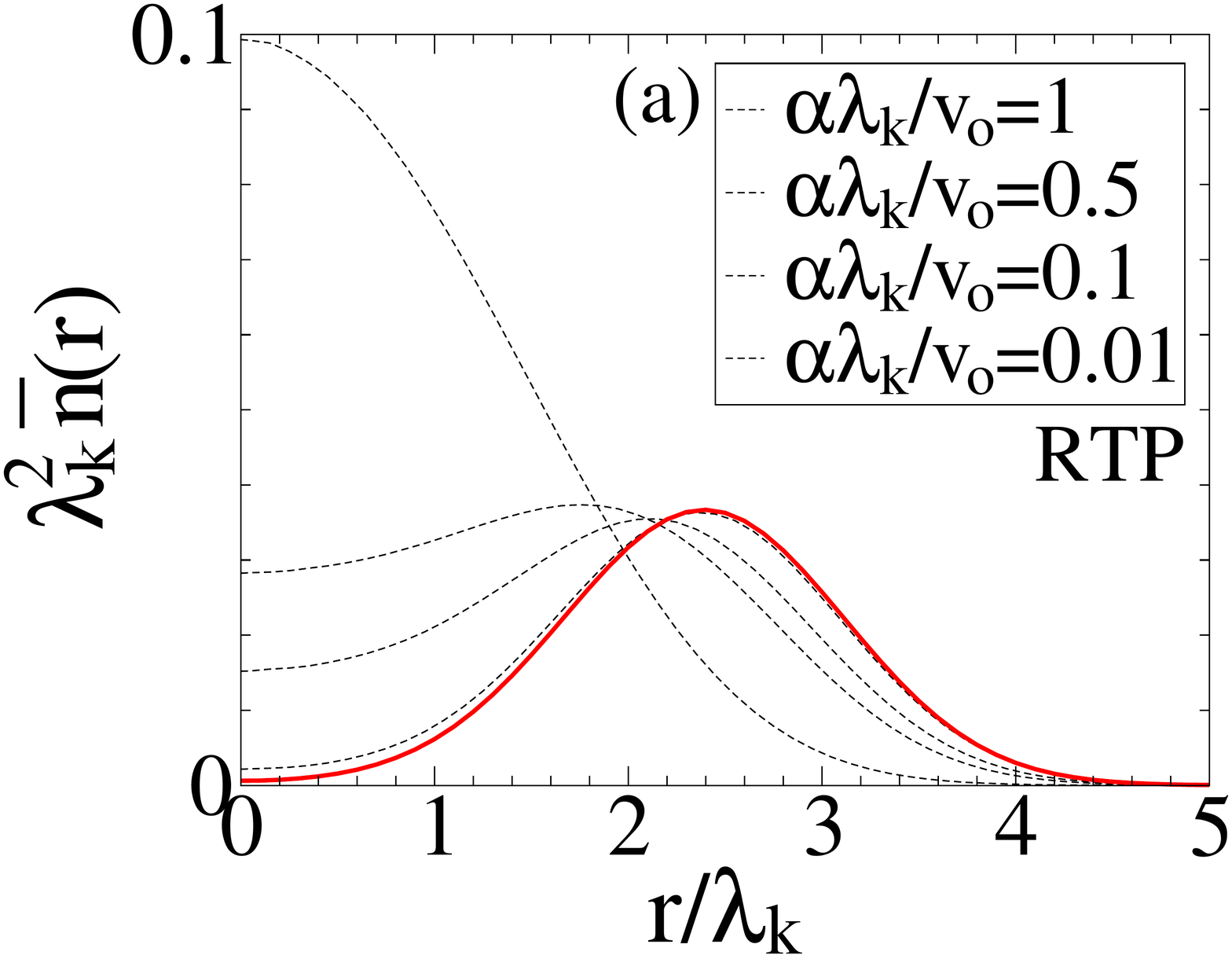} &&
\includegraphics[height=0.21\textwidth,width=0.23\textwidth]{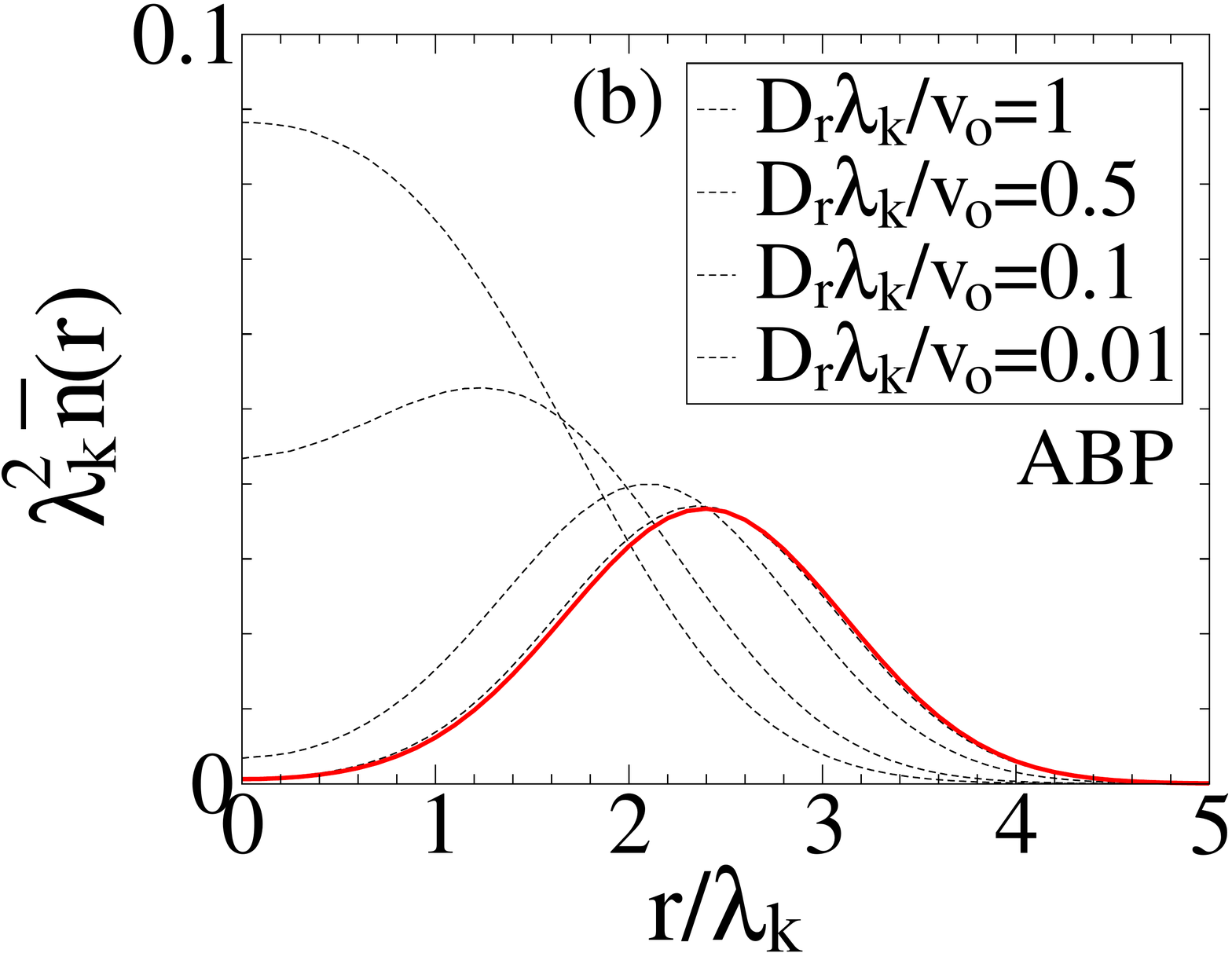}
 \end{tabular}
 \end{center} 
\caption{Distributions of propelled particles in the potential $u_{ext} = {Kr^2}/{2}$ 
obtained from dynamic simulations for $d=2$ (dashed black lines).  $\lambda_k=\sqrt{ {2}/{\beta K}}$ 
is the trap size and the results are for $v_0\lambda_k/D =5$.  The solid red line corresponds to the 
expression in (\ref{eq:n02D}).
The results in (a) are for RTP and those in (b) for ABP type of motion. }
\label{fig:fig1} 
\end{figure}

For dimension $d=3$, the drift orientation is uniformly distributed on a unit sphere with 
$d\Omega_v=\sin\theta_v d\theta_v d\phi_v$.  The disorder averaged distribution obtained 
using Eq. (\ref{eq:n0_ave}) is  
$$
\bar n_0(r) \propto  e^{-\frac{1}{2} \beta Kr^2}   \int_0^{2\pi}d\phi_v \int_0^{\pi} d\theta_v\, \sin\theta_v \,e^{r\cos\theta_v v_0/D},
$$
and evaluates to 
\be
\bar n_0(r)  =  
\bigg[\bigg(\frac{\beta K}{2\pi}\bigg)^{3/2} \  e^{-\frac{1}{2} \beta K r^2} \bigg]  \bigg[ e^{-\frac{v_0^2}{2D^2 \beta K}} \frac{D}{v_0r} \sinh\left(\frac{v_0 r}{D} \right)\bigg].  
\label{eq:n03D}
\ee
The result is similar to that in Eq. (\ref{eq:n02D}).  
The deposition of particles predicted by (\ref{eq:n02D}) and (\ref{eq:n03D}) correspond to the 
optimal deposition.  Any finite value of $\alpha>0$ or $D_r>0$ would make this deposition
less extreme.  To see how the true stationary distributions $\bar n(r)$ evolve toward $\bar n_0(r)$ 
as $\alpha$ or $D_r$ tend to zero, in Fig. (\ref{fig:fig1}) we plot the distributions obtained 
from dynamic simulations for both the RTP and ABP type of motion for particles trapped in the harmonic 
potential and for the dimension $d=2$.  The results are compared to the limiting functional form in Eq. (\ref{eq:n02D}).  
\graphicspath{{figures/}}


\subsection{particles in a confinement with 1D geometry}

If a confining potential has 1D geometry, the system is effectively one-dimensional.  
The simplest example is for particles trapped between two parallel walls.  Since $u_{ext}=0$, 
the effective potential is 
$$
\beta u_{eff}(x) = - \frac{v_xx}{D}, 
$$
where $x$-axis is perpendicular to the walls.  

The normalized Boltzmann distribution for this effective potential, representing the decoupled limit, is  
\be
n_0(x,v_x) = \frac{1}{2h} \frac{v_x h}{D} \frac{e^{\frac{v_xx}{D}}}{\sinh(\frac{v_x h}{D})}.  
\label{eq:n0_wall}
\ee
For the dimension $d=2$ the disorder averaged distribution is given by 
\be
\bar n_0(x) = \frac{1}{2\pi} \int_0^{2\pi} d\theta_v \, n_0(x,v_0\cos\theta_v).  
\label{eq:n02D_wall}
\ee
Using $\frac{d\cos\theta_v}{d\theta_v} = -\sin\theta_v$ and $v_x = v_0 \cos\theta_v$, 
we obtain 
$$
d\theta_v = -\frac{dv_x}{ \sqrt{v_0^2 - v_x^2} }, 
$$
and the integral in Eq. (\ref{eq:n0_wall}) can be rewritten as 
\be
\bar n_0(x) = \frac{1}{\pi} \int_{-v_0}^{v_0} d v_x \, \frac{n_0(x,v_x)}{ \sqrt{v_0^2 - v_x^2} }.  
\ee
Or more generally, we can write 
\be
\bar n_0(x) = \int_{-v_0}^{v_0} d v\,  P(v) n_0(x,v), 
\label{eq:nbar}
\ee
where we use $v \equiv v_x$ and for $d=2$ the distribution  of drifts is 
\be
P(v) = \frac{1}{\pi} \frac{1}{\sqrt{v_0^2 - v^2}}.  
\label{eq:P2}
\ee
Even if the drift orientations 
are uniformly distributed in the variable $\theta_v$, when considering the variable $v_x$, 
there is a considerable inhomogeneity with peaks at $v=\pm v_0$.

For the dimension $d=3$ the disorder averaged distribution is given by 
\be
\bar n_0(x) = \frac{1}{4\pi} \int_0^{\pi} d\theta_v  \int_0^{2\pi} d\phi_v \, \sin\theta_v \, n_0(x,v_0\cos\theta_v), 
\label{eq:n03D_wall}
\ee
and evaluates to (see Appendix \ref{app:1} for the derivation)
$$
\bar n_0(x) =  \frac{1}{2v_0} \int_{-v_0}^{v_0} d v\,  n_0(x,v).  
$$
Comparing to Eq. (\ref{eq:nbar}), this implies that $P(v)$ is uniform on the interval $-v_0 \le v \le v_0$, 
\be
P(v) = \frac{1}{2v_0}.   
\label{eq:P3}
\ee
The expressions in (\ref{eq:P2}) and (\ref{eq:P3}) show strong dependence of $P(v)$ on the system 
dimensionality and suggest that the particle deposition at the walls is larger for $d=2$ than that for $d=3$.  

The next question is, can the integral in (\ref{eq:nbar}) be evaluated exactly.  
Even for uniform distribution $P(v)$, representing the system in $d=3$, 
resulting analytical expression is rather complex.  It involves Hurwitz-Lerch zeta 
and hypergeometric functions.  From practical point of view, it is more convenient to 
evaluate Eq. (\ref{eq:nbar}) numerically for both $d=2$ and $d=3$.



\begin{figure}[h] 
 \begin{center}
 \begin{tabular}{rrrr}
\includegraphics[height=0.21\textwidth,width=0.23\textwidth]{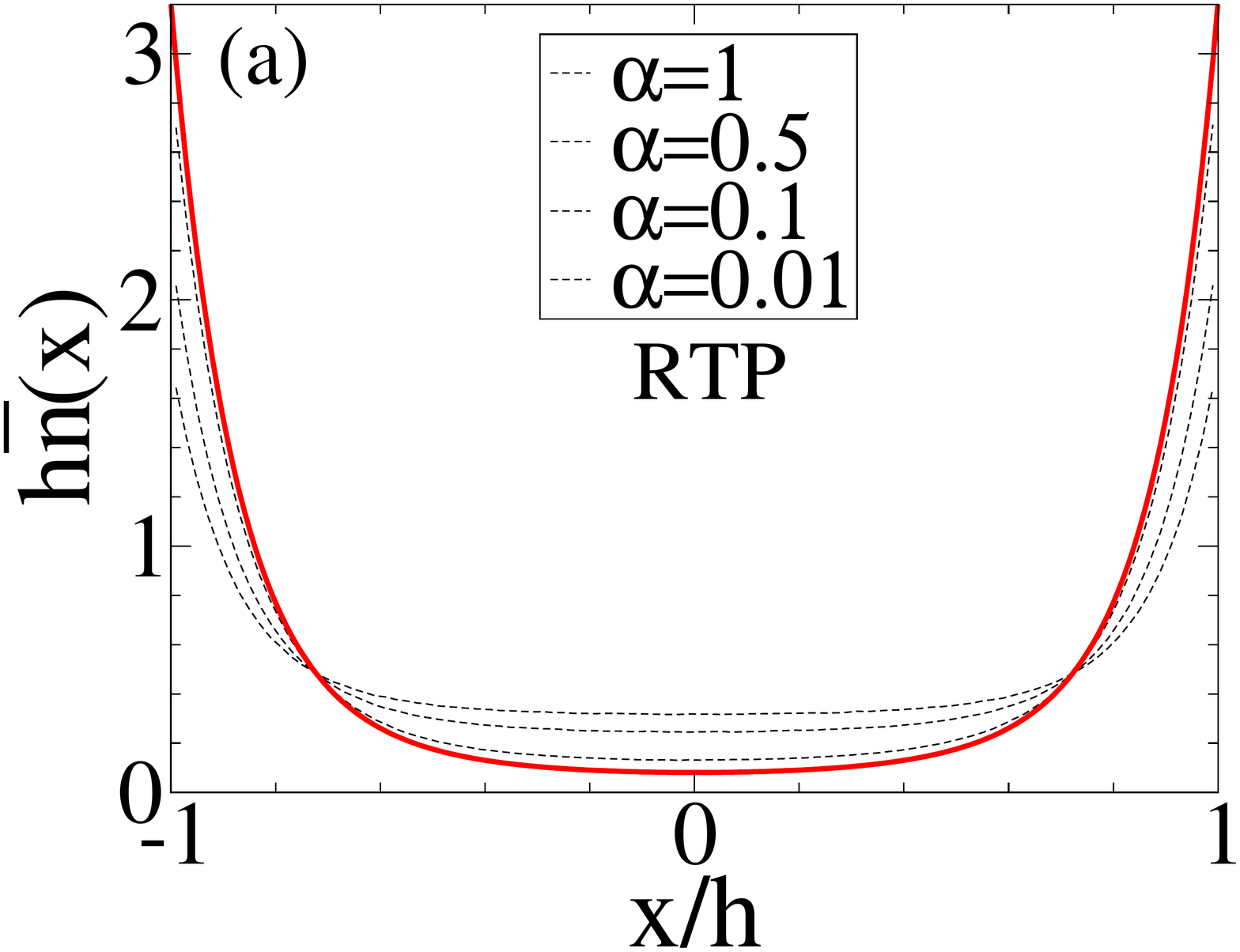} &&
\includegraphics[height=0.21\textwidth,width=0.23\textwidth]{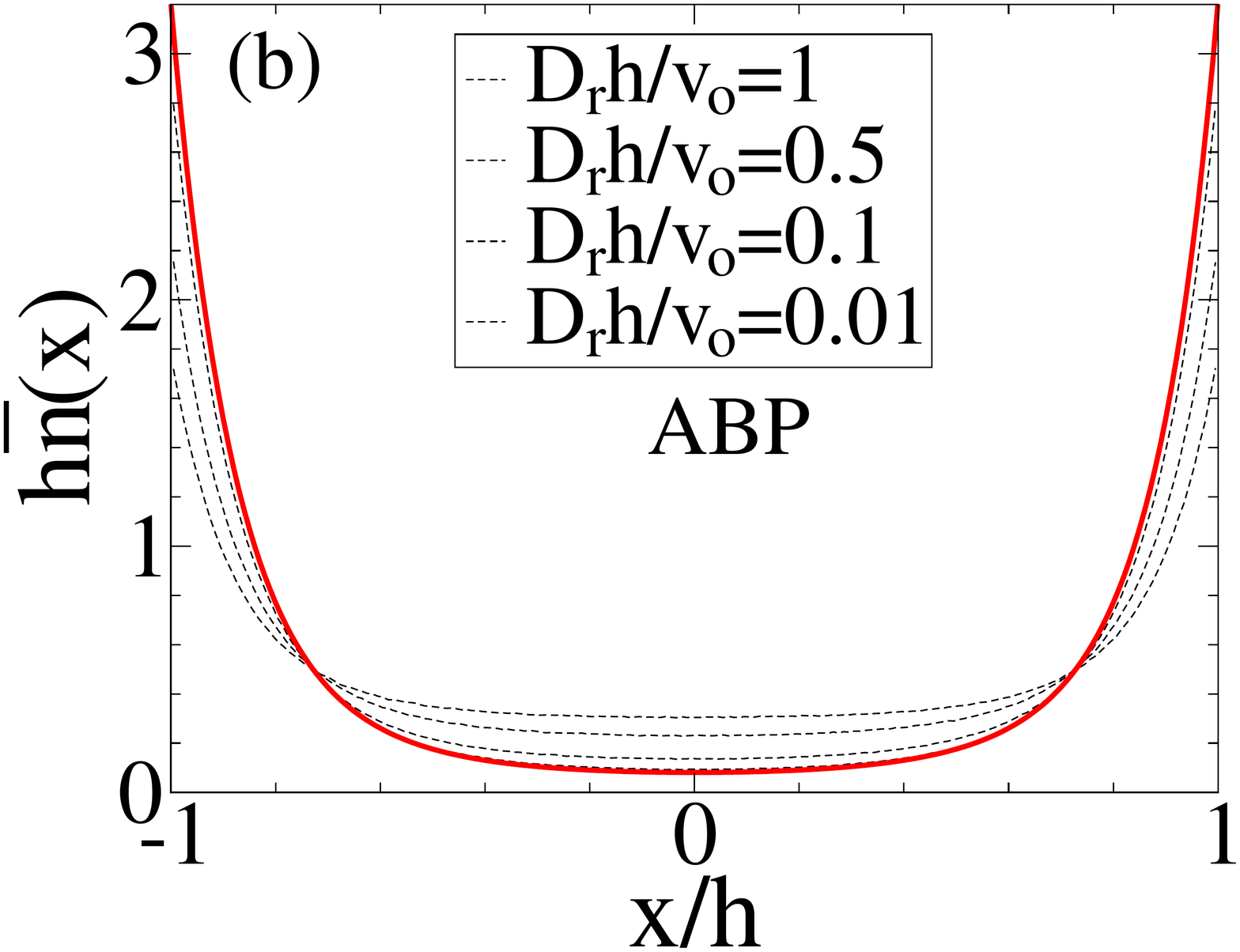}
 \end{tabular}
 \end{center} 
\caption{Distributions of propelled particles between two parallel walls separated by $2h$
obtained from dynamic simulations for $d=2$ (dashed black lines).  
The results are for $v_0 h/D=10$.  
The solid red line corresponds to the distribution $\bar n_0(x)$ in the quenched 
disorder limit.  The results in (a) are for RTP motion and (b) for ABP motion.}
\label{fig:fig1b} 
\end{figure}

In Fig. (\ref{fig:fig1b}) we show an analogous plot to that in (\ref{fig:fig1}) but for 
particles between two parallel walls and decreasing values of $\alpha$
and $D_r$ in order to demonstrate convergence of the distributions to $\bar n_0$.  The 
distribution $\bar n_0$ correctly delimits the range within which the distributions 
$n$ evolve.  
\graphicspath{{figures/}}

A different example of a potential with 1D geometry is the harmonic potential $u_{ext} = \frac{Kx^2}{2}$.  
The normalized Boltzmann distribution corresponding to the decoupled limit in this case is 
$$
n_0(x,v) = \sqrt{\frac{\beta K}{2\pi}} e^{-v^2/2\beta K D^2}e^{\frac{vx}{D} - \frac{\beta Kx^2}{2}}, 
$$
and the disorder averaged distribution is obtained from Eq. (\ref{eq:nbar}) for an appropriate
$P(v)$.  For $d=2$ the integral must be evaluated numerically, and for $d=3$
it evaluates to the following expression 
\be
\bar n_0(x) =
\frac{D}{2v_0 \lambda_k^2}  \bigg( 
\text{erf}\bigg[\frac{x}{\lambda_k} 
 + \frac{1}{2}\frac{v_0 \lambda_k}{D} \bigg] 
-\text{erf}\bigg[\frac{x}{\lambda_k} - \frac{1}{2}\frac{v_0 \lambda_k}{D} \bigg]
\bigg),
\ee 
where $\text{erf}(x)$ is the error function and $\lambda_k=\sqrt{2/\beta K}$.  
Unlike the results in (\ref{eq:n02D}) and (\ref{eq:n03D}), the simple separation between 
the passive and propelled motion is not possible.

\begin{figure}[h] 
 \begin{center}
 \begin{tabular}{rrrr}
\includegraphics[height=0.21\textwidth,width=0.23\textwidth]{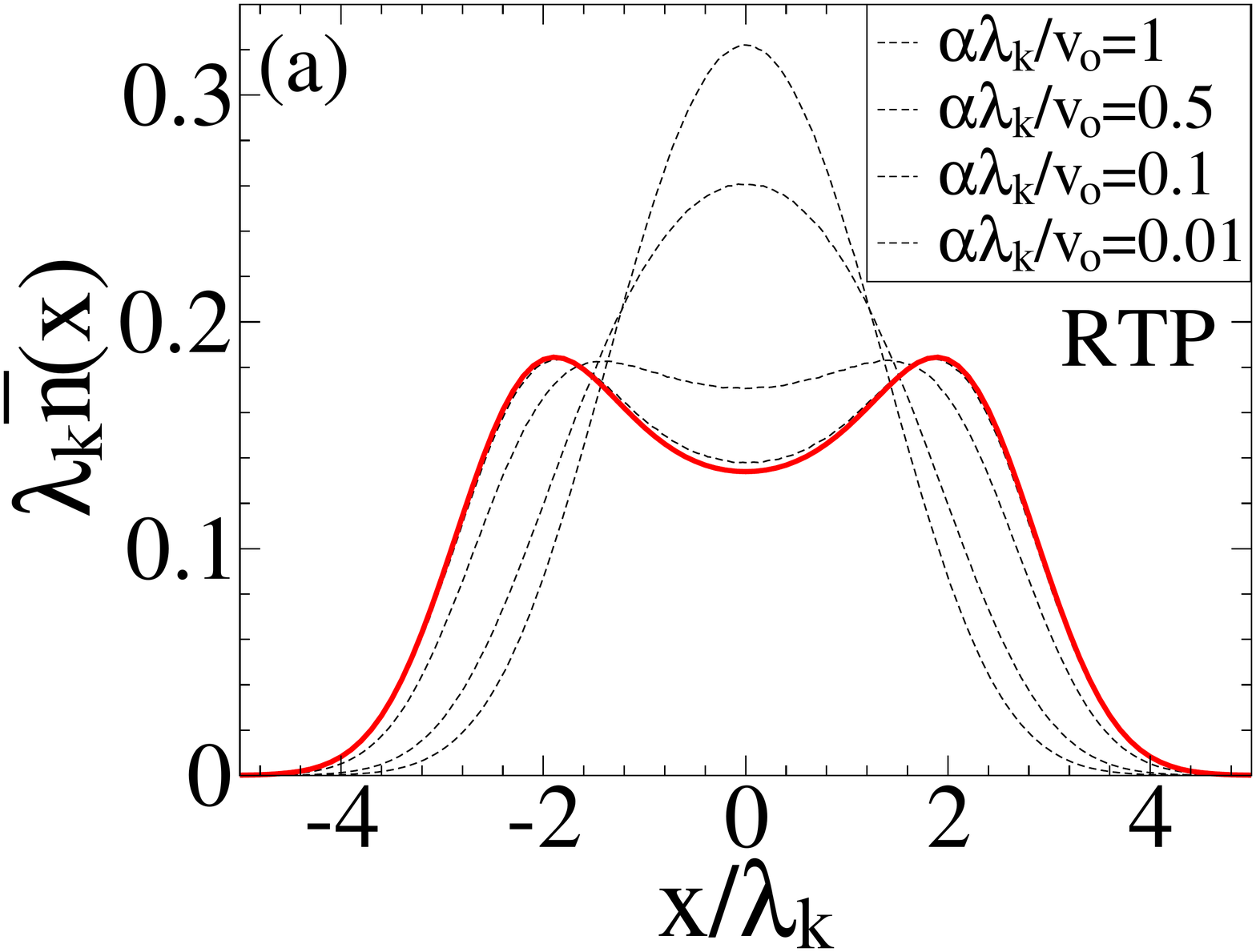} &&
\includegraphics[height=0.21\textwidth,width=0.23\textwidth]{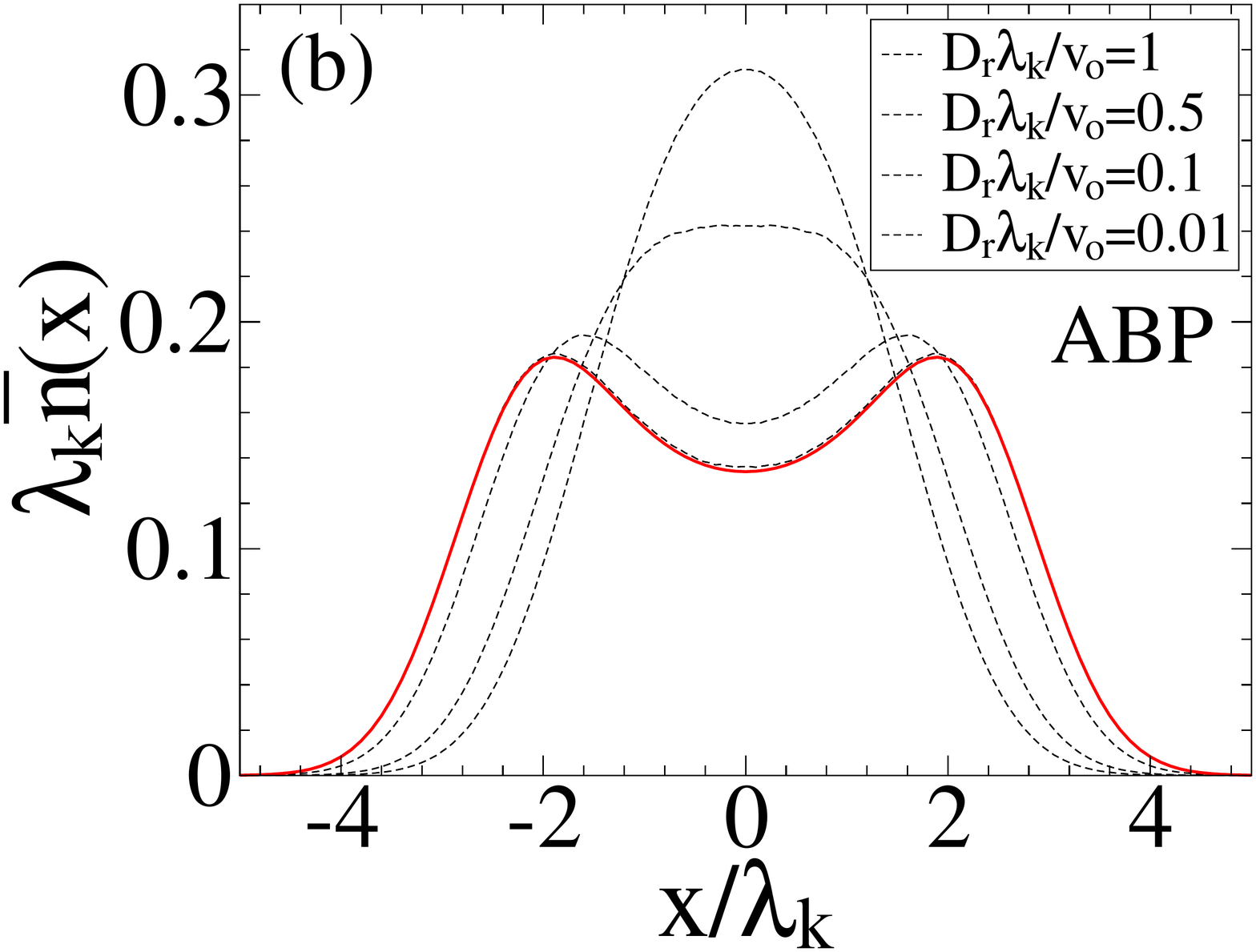}
 \end{tabular}
 \end{center} 
\caption{Distributions of propelled particles in the external potential $u_{ext} = {Kx^2}/{2}$ 
obtained from dynamic simulations for $d=2$ (dashed black lines).  
$\lambda_k=\sqrt{ {2}/{\beta K}}$ is the trap size and the results are for $v_0\lambda_k/D=5$.  
The solid red line corresponds to the distribution $\bar n_0(x)$.  The results in (a) are for RTP 
and (b) for ABP motion. }
\label{fig:fig2a} 
\end{figure}

In Fig. (\ref{fig:fig2a}) we plot the distributions $\bar n(x)$ for 
the potential $u_{ext} = \frac{Kx^2}{2}$ for $d=2$ for decreasing values of 
$\alpha$ and $D_r$, in analogy to figures in (\ref{fig:fig1}) and (\ref{fig:fig1b}).  
Once again, the distributions $\bar n_0$ correctly delimit the range within which 
the true distributions for finite $\alpha$ or $D_r$ can be found.
\graphicspath{{figures/}}

Earlier we briefly discussed the dependence of $P(v)$ on dimensionality when comparing $P(v)$ for $d=2$ 
and $d=3$ in (\ref{eq:P2}) and (\ref{eq:P3}) and the implication of those differences on the accumulation 
of particles at the trap borders.  Below we provide a general expression of $P(v)$, derived in Appendix \ref{app:1}, 
for a general dimension $d>1$, 
\be
P(v) = \frac{1}{v_0\sqrt{\pi}}
\frac{\Gamma[d/2]}{\Gamma[(d-1)/2]} \bigg(1 - \frac{v^2}{v_0^2}\bigg)^{(d-3)/2}, ~~\text{if}~~ d>1, 
\label{eq:Pd}
\ee
with $P(v)$ normalized and defined on the interval $-v_0 \le v \le v_0$.  For $d=1$ the distribution is 
represented in terms of delta functions as drifts in this dimension are limited to two values 
$v=\pm v_0$ \cite{Razin20}, 
\be
P(v) =  \frac{1}{2}\big[\delta(v+v_0)+\delta(v-v_0)\big], ~~\text{if}~~ d=1
\label{eq:P1}
\ee
Clearly, the distribution $\bar n_0$ calculated using (\ref{eq:nbar}) depends on $P(v)$.  
For large $d$ the distribution $P(v)$ approaches a Gaussian functional form 
$$
P(v) \approx \sqrt{\frac{d}{2\pi v_0^2}} e^{-\frac{d}{2}(v/v_0)^2},
$$
and in the limit $d\to\infty$, $P(v)\to \delta(v)$, and the system loses its quenched
disorder --- all particles have zero drift and the system becomes identical with that for passive Brownian
particles. 

\begin{figure}[h] 
 \begin{center}
 \begin{tabular}{rrrr}
\includegraphics[height=0.21\textwidth,width=0.23\textwidth]{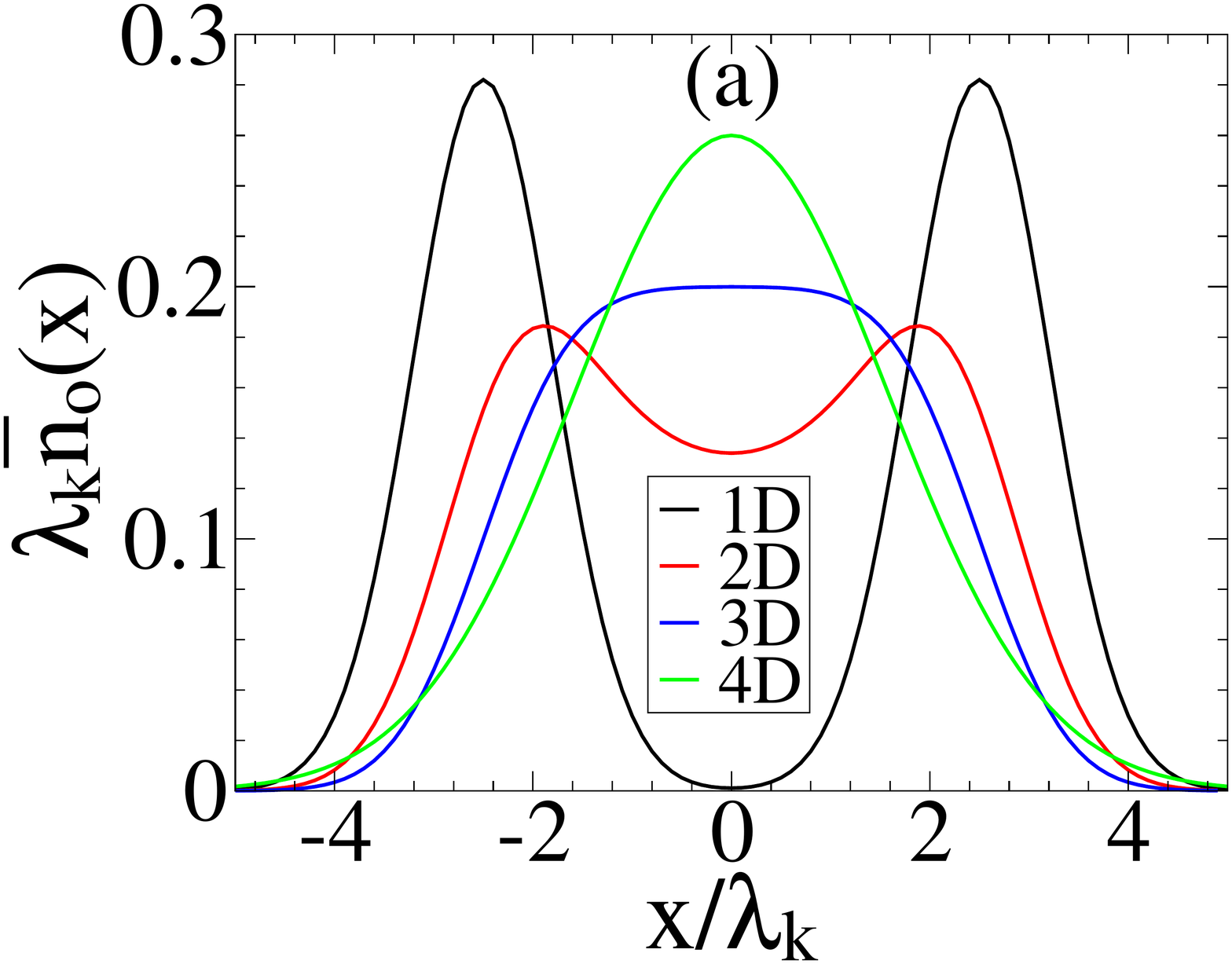}&
\includegraphics[height=0.21\textwidth,width=0.23\textwidth]{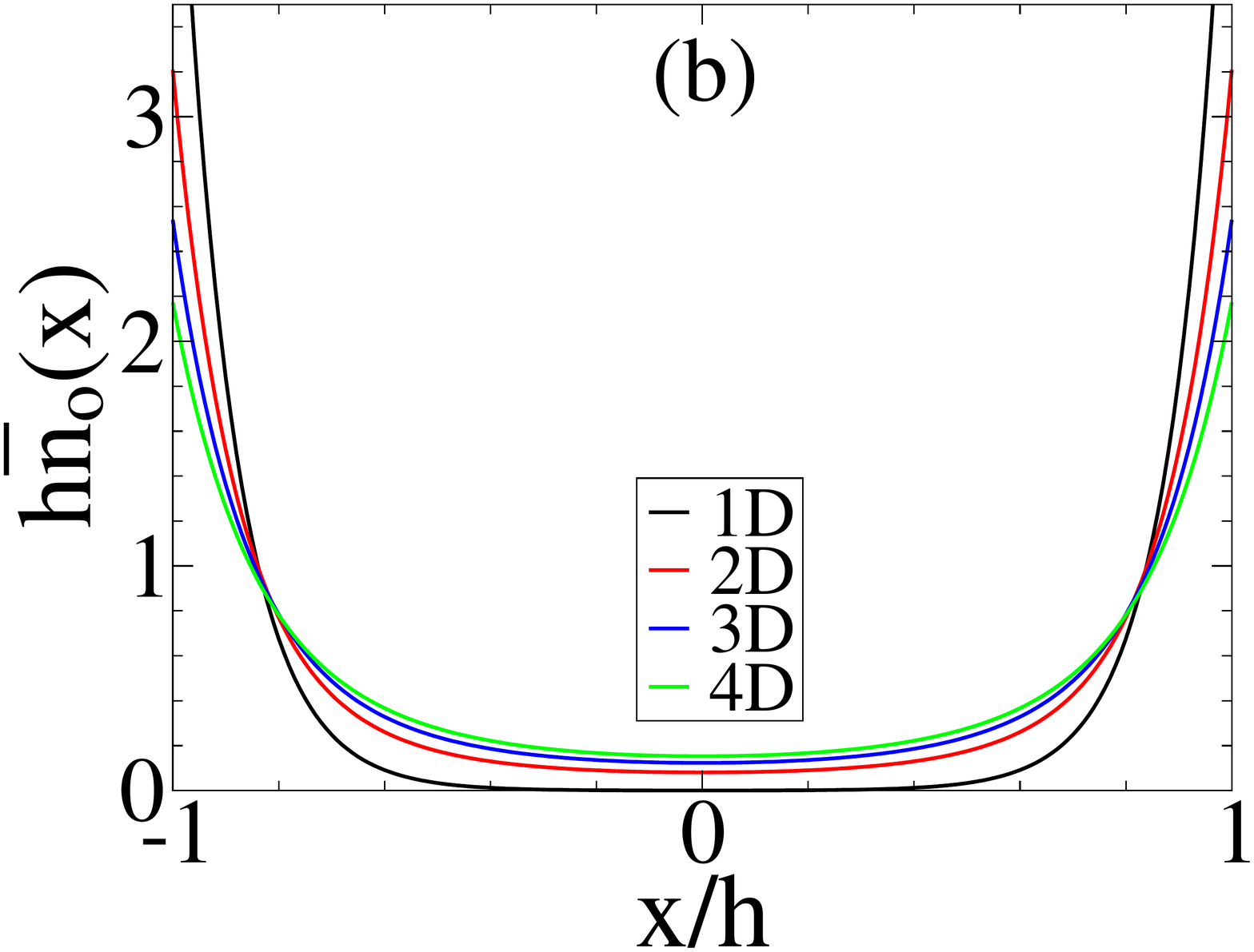}\\
 \end{tabular}
 \end{center}
\caption{Distributions $\bar n_0(x)$ for an external potential (a) $u_{ext}=Kx^2/2$ 
and (b) for particles between two walls, for different system dimensionality $d$.  
The distributions are for $v_0 \lambda_k/D=5$ and $v_0 h/D=10$.}
\label{fig:fig2b} 
\end{figure}

In Fig. (\ref{fig:fig2b}) we plot the distributions $\bar n_0(x)$ for two different external traps, 
$u_{ext} = \frac{Kx^2}{2}$ and for confinement between two walls, for $P(v)$ in (\ref{eq:Pd})
and (\ref{eq:P1}) corresponding to different $d$.  The plots demonstrate strong dependence 
on $d$, in particular, it shows increased deposition of particles around the trap borders as 
dimensionality goes down.  
\graphicspath{{figures/}}

A similar dimensionality dependence is found 
in the opposite limit of large $\alpha$ and/or $D_r$, accurately represented by the concept of effective 
temperature \cite{Kardar15,Kurchan13,Szamel14} valid for $d>1$, 
\be
\frac{T_{\rm eff}}{T} = 1 + \frac{1}{d(d-1)} \frac{v^2_0}{D(\alpha + D_r)},
\label{eq:Teff}
\ee
where increased dimensionality $d$ brings $T_{eff}$ closer to thermodynamic temperature $T$.  
In the limit $d\to \infty$, $T_{eff}=T$.  The reason for this 
behavior is rather simple.  The constant velocity $v_0$ and the associated kinetic energy is distributed into $d$  
components.  For increased dimensionality, the extra kinetic energy that goes to each degree of freedom
is reduced, giving rise to the observed cooling effect.

\section{Self-consistent formulation}
\label{sec:SC}

The next step is to try to expand the distribution $n$ around the decoupled limit as 
$n \approx n_0 + \alpha n_1$.  However, such a systematic expansion yields 
expressions which are complex and not very insightful.  Instead, we reformulate the 
stationary FP equation as a self-consistent relation (SC).  The resulting formulation 
yields interesting insights, provides basis for an alternative computation of distributions, 
and can be used for obtaining perturbative expansion of $n$.

\subsection{RTP particles}

To keep things simple, we consider a system with 1D geometry.  For the RTP motion the 
stationary FP equation in 1D can be written as 
$$
0 = Dn''  + \beta D \big[u_{eff}' n\big]'  + \alpha (\bar n -  n), 
$$
where the effective potential incorporates the drift as 
$
\beta u_{eff} = \beta u_{ext} - \frac{v}{D}x, 
$
and the disorder averaged distribution is $\bar n = \int dv\, P(v) n(n,x)$.  
The same equation can be written as 
\be
0 = n''  + \beta \big[u_{eff}' n \big]'  +   s
\label{eq:FPS2}
\ee
where
\be 
s = \frac{\alpha}{D} (\bar n - n)
\label{eq:source}
\ee 
plays the role of the source function.  Note that the source function satisfies 
$\int dx\,s(x,v) = 0$ and $\int dv\, P(v) s(x,v) = 0$.  

By introducing the 
source function, Eq. (\ref{eq:FPS}) can be regarded as an inhomogeneous second-order 
differential equation.  The solution then can be obtained using the method of variation of parameters.  
To proceed, we first need solutions for the homogenous equation.  The two possible
solutions are 
\be
y_0 = e^{-\beta u_{eff}}, ~~~~ y_1 = y_0 Y_0,
\label{eq:y0}
\ee
where 
\be
Y_0 =  \int dx\, e^{\beta u_{eff}}.  
\ee
The first solution corresponds to the Boltzmann distribution.  The second solution is normally 
rejected on physical grounds due to its non-vanishing local flux, $D\rho' + u_{eff}'\rho\ne 0$, 
when dealing with passive particles.  
As we will see, this solution becomes relevant for describing propelled particles.  

The solution of the second order inhomogeneous equation 
can be expressed as  
\be
n = A y_0 + B y_1 + \bigg[y_0 \int dx\, \frac{y_1}{w} s  -  y_1 \int dx\, \frac{y_0}{w}  s\bigg], 
\ee
where $A$ and $B$ are undefined coefficients and 
$w = y_0y_1' - y_0'y_1$  is the Wronskian that for the present case evaluates as $w = y_0$.  
The first two terms constitute a complementary solution 
and the last term is the particular solution.  Since the second term does not produce a vanishing flux, 
$B$ is set to zero.  After using (\ref{eq:y0}) and substituting (\ref{eq:source}) for the source
function, the solution transforms into the desired SC relation 
\be
n  = A e^{-\beta u_{eff}}  + \frac{\alpha e^{-\beta u_{eff}}}{D} \bigg[\int dx\,  (\bar n-n)Y_0 - Y_0 \int dx\, (\bar n-n)  \bigg], 
\label{eq:n_sc}
\ee
where $A$ is determined from the condition of normalization
$\int_{L} dx\,n(x,v) = 1$ on the domain $L$ prescribed by a physical problem.  Note that 
for $\alpha=0$, we recover $n=n_0$.  

The SC relation in (\ref{eq:n_sc}) reveals a certain mean-field character of the formulation 
\cite{Frydel16} and the presence of the effective interactions between particles --- particles 
appear to be "attracted" toward the average distribution $\bar n$.  The origin of this coupling 
between particles, however, is different from that in a system of truly interacting particles.  
It is caused by the "reaction" part of the FP equation, as particles of different drift, regarded
as belonging to different species, exchange their identity.

If the RTP particles are confined between two parallel walls then 
$
\beta u_{eff} = -\frac{vx}{D}, 
$
and
$$
Y_0 = -\frac{D}{v} e^{-\frac{vx}{D}}, 
$$
and the SC relation becomes  
\be
n  = A n_0  + \alpha \int_{-h}^{x} dx'\,  \bigg[\frac{1-e^{\frac{v}{D} (x-x')}}{v} \bigg] (\bar n  - n).  
\label{eq:sc_wall}
\ee

The above SC relation is next used as a basis for numerical computation of the distributions $n$
based on iterative procedure starting with 
$n_0$.  For $\alpha \ge 0.5$ a mixing parameter is used, $0< \gamma < 1$, for generating the
next distribution as $n_{new} \equiv (1-\gamma) n_{old} + \gamma n_{new}$.  For the bin size 
$\Delta x = 0.01$ the convergence is attained within ten to twenty iterations
(amounting to a few seconds of the CPU time, a significant improvement over dynamic simulations).

\begin{figure}[h] 
 \begin{center}
 \begin{tabular}{rrrr}
\includegraphics[height=0.21\textwidth,width=0.25\textwidth]{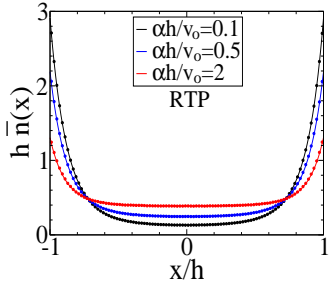}\\
 \end{tabular}
 \end{center}
\caption{Distributions $\bar n(x)$ obtained numerically using the SC formulation of the 
FP equation for the RTP particles between two walls 
for $d=2$ and $v_0h/D=10$.  Circles correspond to simulation data points. }
\label{fig:fig3} 
\end{figure}

Fig. (\ref{fig:fig3}) plots the numerically calculated stationary distributions  
for $d=2$ (using the distribution $P(v)$ in (\ref{eq:P2})).  The distributions are 
in perfect correspondence with those obtained from dynamic simulations.  
\graphicspath{{figures/}}

The SC formulation in (\ref{eq:n_sc}), or that for particles between walls in (\ref{eq:sc_wall}), can 
also be used for constructing subsequent terms within the perturbative approach, 
$n = n_0 + \alpha n_1 + \dots$, by inserting 
$n_0$ on the right hand side of those equations.  If considering Eq. (\ref{eq:sc_wall}), we get 
\be
n_1 =  \int_{-h}^{x} dx'\,  \bigg[\frac{1-e^{\frac{v}{D} (x-x')}}{v} \bigg] (\bar n_0  - n_0) + C n_0,
\label{eq:n1}
\ee
where the constant $C$ is such as to ensure the condition $\int_{-h}^h dx\, n_1(x,v)=0$, since the perturbation
$n_1$ cannot create or destroy particles, only redistribute them in the interval $-h\le x \le h$.  
We recall that $n_0(x,v)$ for the system between walls is given in Eq. (\ref{eq:n02D_wall}),   
however, inserting this expression into (\ref{eq:n1}) does not lead to analytical results and the perturbative 
formulation itself does not shed any additional light.

\begin{figure}[h] 
 \begin{center}
 \begin{tabular}{rrrr}
\includegraphics[height=0.21\textwidth,width=0.25\textwidth]{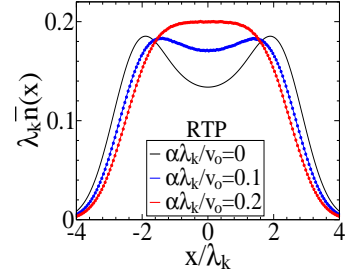}\\
 \end{tabular}
 \end{center}
\caption{Distributions $\bar n(x)$ obtained numerically using the SC formulation for the RTP particles 
 in the potential $u_{ext} = Kx^2/2$ for $d=2$ and $v_0\lambda_k/D=5$.  Circles correspond to 
 simulation data points.  }
\label{fig:fig4} 
\end{figure}

We next consider a harmonic potential, in which case 
$
\beta u_{eff} =  - \frac{vx}{D}  + \frac{\beta Kx^2}{2}, 
$
$$
Y_0 = \frac{\lambda_k\sqrt{\pi}}{2} e^{-(\frac{v\lambda_k}{2D})^2}  \text{erfi}\Big[\frac{x}{\lambda_k} - \frac{1}{2}\frac{v\lambda_k}{D}\Big], 
$$
where $\text{erfi}(x) = -i \text{erf}(ix)$ is the imaginary error function,  
and the SC equation expressed in terms of definite integrals is
\be
n e^{\beta u_{eff}} = A  + \frac{\alpha}{D} \bigg[\int_{-\infty}^x dx'\,  (\bar n-n)Y_0 - Y_0 \int_{-\infty}^x dx'\, (\bar n-n)  \bigg].
\label{eq:n_sc2}
\ee
For numerical integration the limits $x=\pm\infty$ are substituted by $x=\pm x_c$ where the cutoff distance $x_c$ 
is large enough so that $n(\pm x_c,v)\approx 0$.  
Numerically calculated distributions for $d=2$ are shown in Fig. (\ref{fig:fig4}).  
\graphicspath{{figures/}}
Again, the distributions are in perfect correspondence with those obtained from dynamic simulations.

\subsection{ABP particles}

Self-consistent relation could similarly be established for the ABP type of motion.  Considering 
the system dimension $d=2$ and a system with 1D geometry, the stationary FP equation that
describes this situation, obtained using Eq. (\ref{eq:FP2D}) with $\alpha=0$ but for finite $D_r$, is 
\be
0 = n'' + \beta D \big[u_{eff}' n\big]'  +  \frac{D_r}{D} \frac{\partial^2n}{\partial\theta_v^2},  
\ee
where $n\equiv n(x,\theta_v)$.  If the source term is defined as 
$$
s = \frac{D_r}{D} \frac{\partial^2n}{\partial\theta_v^2},
$$
we arrive at a similar form to that in (\ref{eq:FPS2}), and can follow up with the same
procedure.    
In the case of ABP motion, the expressions are more economic if the distributions are
defined in terms of $\theta_v$ rather than $v\equiv v_x$.

\begin{figure}[h] 
 \begin{center}
 \begin{tabular}{rrrr}
\includegraphics[height=0.21\textwidth,width=0.23\textwidth]{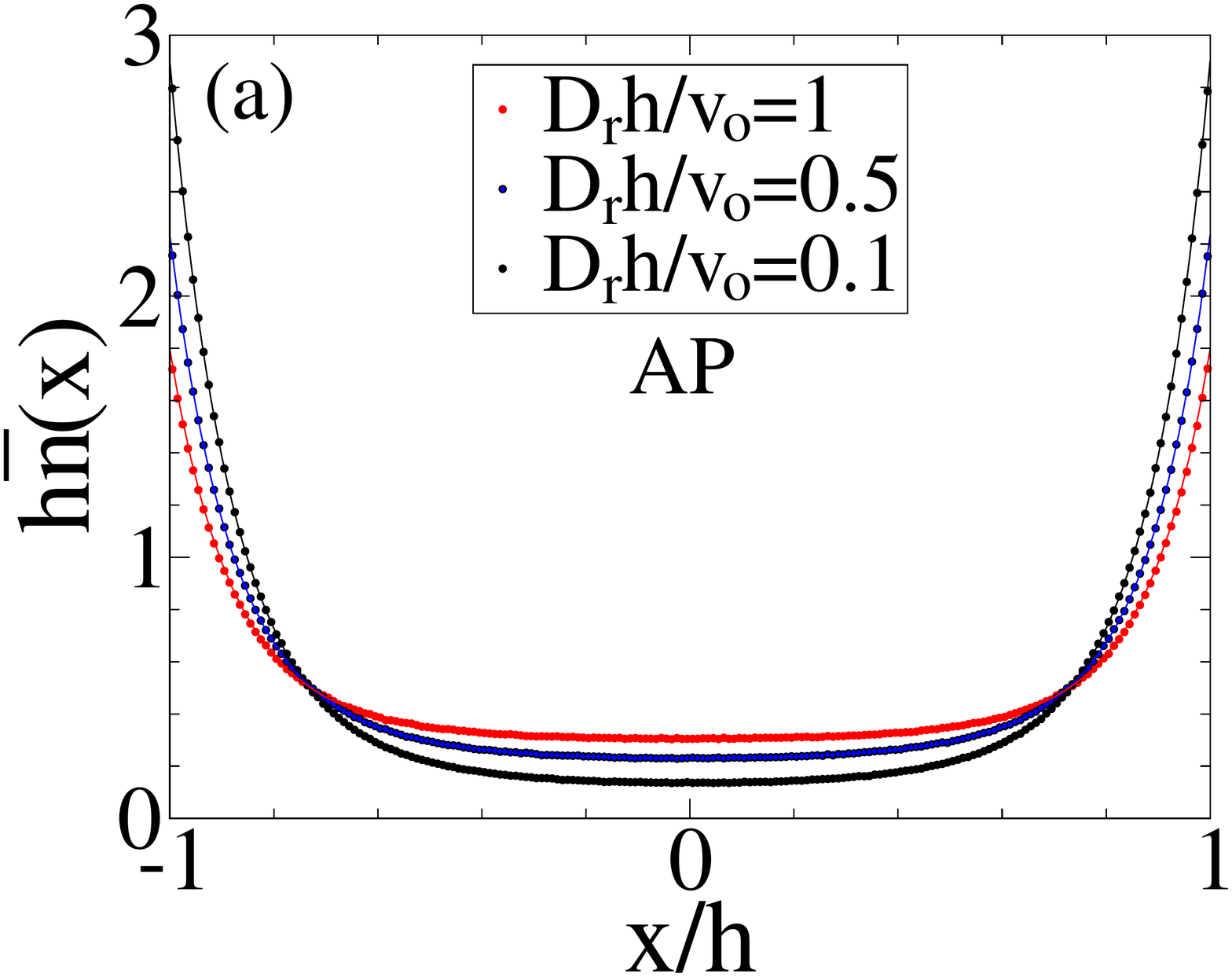}&
\includegraphics[height=0.21\textwidth,width=0.23\textwidth]{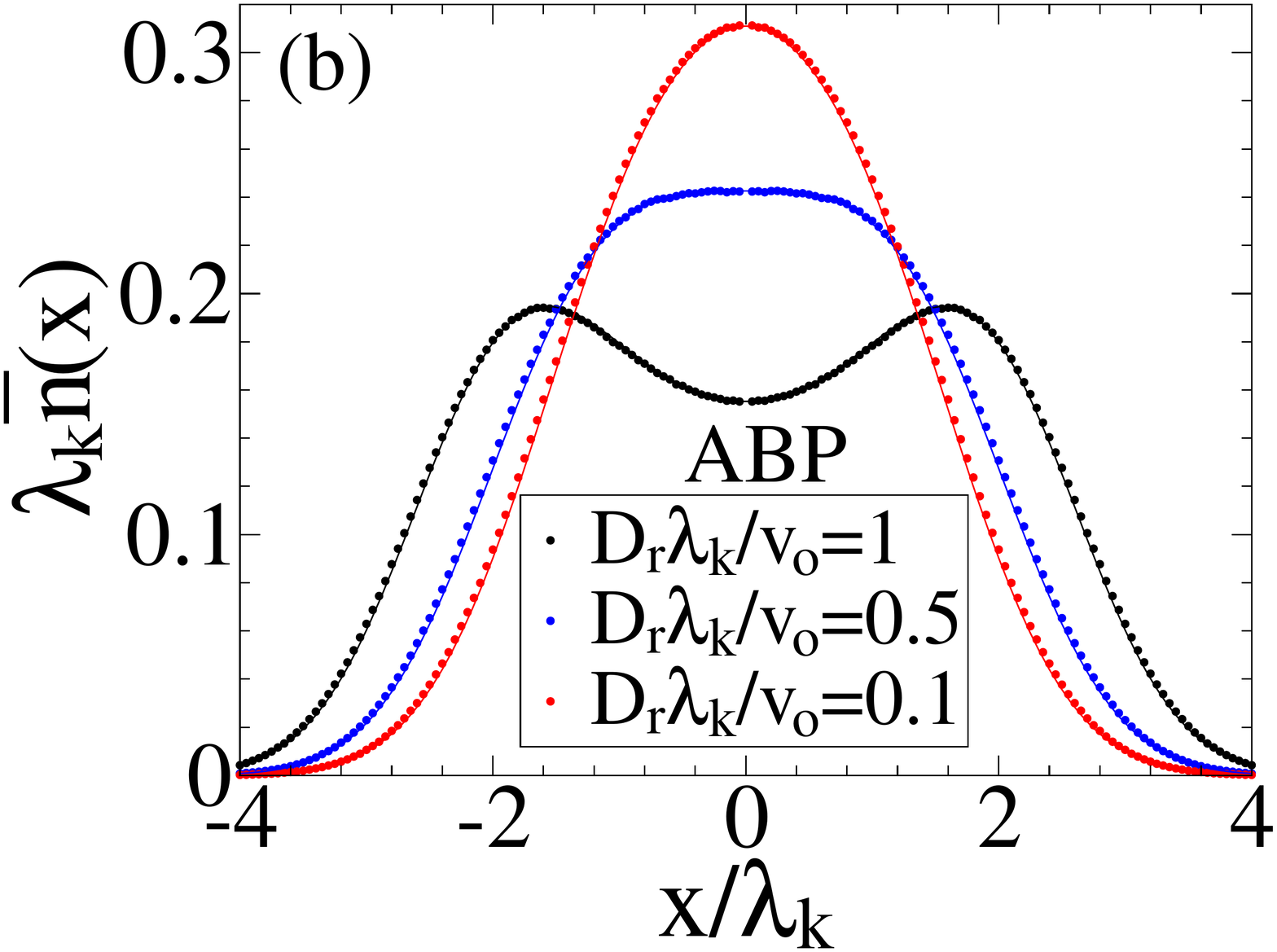}\\
 \end{tabular}
 \end{center}
\caption{Distributions $\bar n(x)$ obtained numerically from the SC formulation for the ABP particles 
for $d=2$ (a) between two walls with $v_0h/D=10$ and (b) in the harmonic potential with $v_0\lambda_k/D=5$.  
Circles correspond to simulation data points. }
\label{fig:fig3a} 
\end{figure}

The SC relation that follows is 
\be
n  = Ae^{-\beta u_{eff}}  
+ \frac{D_re^{-\beta u_{eff}}}{D} \bigg[\int dx\,   Y_0\, \frac{\partial^2n}{\partial\theta_v^2}    - Y_0 \int dx\,  \frac{\partial^2n}{\partial\theta_v^2}  \bigg], 
\label{eq:n_sc2}
\ee
and can next be used as a basis for calculating stationary distributions.  
The results are shown in Fig. (\ref{fig:fig3a}).  
Unlike for the RTP particles, the numerical method is less 
robust and larger number of iteration is required to reach convergence.  
\graphicspath{{figures/}}




\section{What is the true equilibrium?}
\label{sec:diss}

There is an interesting consequence of treating the system with $\alpha=D_r=0$
as a reference point and considering deviations from it as a "distance" from equilibrium.  
According to this viewpoint, the system at $\alpha\to\infty$ or $D_r\to \infty$, represents the 
largest deviation --- the conclusion that runs counter to more conventional point of view 
that regards as a reference state (and equilibrium) the limit $\alpha\to\infty$ or $D_r\to \infty$.


One way to resolve this controversy, of which reference point corresponds to equilibrium, 
is to resort to the arbitration of the entropy production, considered as a sophisticated way of 
quantifying the degree of violation of detailed-balance condition.  We will not make calculations 
for the entropy production for our system.  Instead we use the exact expression for the RTP s
ystem in $d=1$, where $P(v)=\frac{1}{2}[\delta(v-v_0) + \delta(v+v_0)]$, found in Ref. \cite{Razin20}
in Eq. (17) and given by 
\be
\Pi = \alpha \frac{h k \cosh hk - \sinh hk}{\frac{\alpha D}{v_0^2} h k \cosh hk + \sinh hk},
\ee
where $k = \frac{v_0}{D}\sqrt{1+\frac{\alpha D}{v_0^2}}$.  In true equilibrium, $\Pi=0$.  
The larger the value of $\Pi$, we larger the deviation from equilibrium.  If we plot 
$\Pi$ as a function of $\alpha$ for fixed $D$ and $v_0$ we discover that $\Pi(\alpha=0)=0$
and as $\alpha$ increases $\Pi$ grows monotonically and 
in the limit $\alpha\to \infty$ we have $\Pi(\alpha\to \infty)=\frac{v_0^2}{D}$.  
Such result appears to vindicate our viewpoint that the "correct" equilibrium corresponds 
to the decoupled limit, not the other way around.  The reason for this surprising result is that 
even if the distribution $n$ becomes flat and the transport due to diffusion vanishes, a convective 
type of motion is still there.

\section{Conclusion}
 
This work starts by recognizing that at the precise condition $\alpha=D_r=0$, where orientation 
of the drifts becomes fixed and time independent, the system attains an equilibrium with quenched 
disorder.  This intuitive interpretation permits us to obtain exact stationary distributions 
of propelled particles in confining potentials.  The central quantity that emerges is the 
effective potential $u_{eff}$, which is the sum of an external potential and a linear potential 
for representing drift, and the Boltzmann factor 
$e^{-\beta u_{eff}}$.  

In the second part of this work we construct the theoretical framework in which the decoupled
state figures naturally.  This is done by reformulating the stationary FP equation as a self-consistent
relation, formulated in terms of the Boltzmann factor $e^{-\beta u_{eff}}$.  The formulation   
reveals the presence of coupling between propelled particles (even if there are no 
true interactions between particles) as a result of "chemical" process, whereby particles
with different drift are represented as different species that continuously exchange 
identities.  The self-consistent formulation is used as a basis for numerical computation 
of stationary distributions, as an alternative to dynamic simulations.  The SC formulation
can also be used to expand $n$ perturbatively around $n_0$.

The viewpoint that considers the decoupled condition as an equilibrium state raises the question, 
so what the real equilibrium is?  Generally, this privileged status is attributed to the limit $\alpha\to\infty$ 
and/or $D_r\to\infty$, since the distribution in those limits converges to that of passive Brownian 
particles.  However, if we look into the entropy production $\Pi$ that is supposed to measure a 
distance from an equilibrium, we get the results that support the case for the decoupled limit
as a true equilibrium.  

The viewpoint that considers the decoupled condition as an equilibrium state raises the question, 
So what is the real equilibrium? Generally, equilibrium is attributed to the limit $\alpha\to\infty$ and/or $D_r\to\infty$,
since the distribution in those limits converges to that of passive Brownian particles. However, if we 
look into the entropy production $\Pi$ that is supposed to measure a distance from equilibrium, we get 
the results that support the case for the decoupled limit as a true equilibrium.

\begin{acknowledgments}
D.F. acknowledges financial support from FONDECYT through grant number 1201192. 
D.F. thanks the University of Tel Aviv for invitation under the program the ''Visiting Scholar 
of The School of Chemistry''.   
\end{acknowledgments}

\section{DATA AVAILABILITY}
The data that support the findings of this study are available from the corresponding author upon 
reasonable request.

\appendix

\section{Distributions $P(v)$ for a general $d$-dimension}
\label{app:1}

A general, disorder averaged distribution over drift orientations uniformly distributed on the 
surface of a unit sphere in $d$-dimension for the system with 1D geometry, such a system 
between two parallel walls or in the harmonic potential $u_{ext} = \frac{Kx^2}{2}$, is 
$$
\bar n_0(x) \propto \int_{0}^{2\pi}   d\Omega_v\, n_0(x,v_0\cos\theta_v),
$$
where $v_x=v_0\cos\theta_v$ is the velocity component in the direction perpendicular to the boundaries 
of a trap.  The stationary distribution is uniform in the remaining directions.

Since for an arbitrary dimension $d$, $d\Omega$ is defined as 
$$
d\Omega =  \sin^{d-2}\varphi_{1} \sin^{d-3}\varphi_{2} \cdots \sin d\varphi _{d-2}\, d\varphi_{1}\,d\varphi _{2}\cdots d\varphi _{d-1}, 
$$
where $\theta=\varphi_{1}$, we may write 
$$
\int d\Omega_v \, n_0(x,v_0\cos\theta_v) \propto  \int_0^{\pi} d\theta_v  \, \sin^{d-2} \theta_v \,\, n_0(x,v_0\cos\theta_v),
$$
as the angles $\varphi_{k}$ for $k>1$ can be ignored.  
The above integral is transformed using
$
d\theta_v  = -\frac{1}{v_0}\frac{dv}{\sin\theta_v},
$
where $v\equiv v_0\cos\theta_v$, and $\sin\theta_v = \sqrt{1 - \cos^2\theta_v}$ into 
$$
 \int d\Omega_v \, n_0(x,v_0\cos\theta_v) \propto \int_{-v_0}^{v_0} d v  \, 
 \bigg(1 - \frac{v^2}{v_0^2}\bigg)^{\frac{d-3}{2}} \, n_0(x,v),
$$
and the normalized distribution $P(v)$ for an arbitrary dimension $d$ is
\be
P(v) = \frac{1}{v_0\sqrt{\pi}}
\frac{\Gamma[\frac{d}{2}]}{\Gamma[\frac{d-1}{2}]} \bigg(1 - \frac{v^2}{v_0^2}\bigg)^{\frac{d-3}{2}}.  
\ee



\end{document}